\documentclass[letter,twocolumn]{jpsj3}

\newcommand{\lsim} 
 {\ \raise.35ex\hbox{$<$}\kern-0.75em\lower.5ex\hbox{$\sim$}\ }
\newcommand{\gsim}
 {\ \raise.35ex\hbox{$>$}\kern-0.75em\lower.5ex\hbox{$\sim$}\ }

\usepackage{graphicx}
\usepackage{dcolumn}
\usepackage{bm}
\usepackage{amsmath}
\usepackage{times}
\usepackage[usenames]{color}

\title{Charge Dynamics in a Correlated Fermion System on a Geometrically Frustrated Lattice}  
\author{Makoto~Naka$^1$ and Sumio Ishihara$^{2,3}$}
\inst{
$^1$RIKEN Center for Emergent Matter Science (CEMS), Wako 351-0198, Japan \\
$^2$Department of Physics, Tohoku University, Sendai 980-8578, Japan \\
$^3$Core Research for Evolutional Science and Technology (CREST), Sendai 980-8578, Japan}
\abst{
Charge dynamics in an interacting fermionic model on a geometrically frustrated lattice are examined. We analyze a spinless fermion model on a paired triangular lattice, an electronic model for layered iron oxides, in zero and finite temperatures by the exact diagonalization methods. We focus on the two charge ordered (CO) phases, termed CO$_{1/2}$ and CO$_{1/3}$, which, respectively, are realized by the inter-site Coulomb interaction and the quantum/thermal fluctuations. The optical spectra in CO$_{1/3}$ show multiple components and their low-energy weights are survived even below the ordering temperature ($T_{\rm CO}$). Changes of the dynamical charge correlation below $T_{\rm CO}$ in CO$_{1/3}$ are weakly momentum dependent, in sharply contrast to CO$_{1/2}$. These characteristic dynamics in CO$_{1/3}$ are attributable to the charge frustration effects, and reproduce some aspects of the recent experimental optical and x-ray sacattering spectra.
}
\kword{Frustration, Charge Order, Charge Dynamics}

\begin{document}
\maketitle
Frustration is one of the fascinating themes in modern solid state physics.~\cite{jpsj,diep} 
A macroscopic number of degenerated classical states, in which all equivalent interaction energies are not minimized simultaneously, often emerge in several systems on geometrically frustrated lattices. 
Localized electron magnets in triangular, Kagom\'{e}, and pyrochlore lattices are the well-known examples. 
A number of exotic phenomena, such as quantum-spin liquid states, nematic spin state, 
multiferroics, and so on have attracted broad interest from the theoretical and experimental view points. 

Interacting itinerant electron system on a geometrical frustrated lattice is another example.~\cite{poilblanc, seo,merino} 
A rich variety of novel phenomena in the frustrated electron systems
has been discovered in broad classes of transition-metal oxides and organic molecular solids, and has been examined in the theoretical calculations. 
Essential points of the frustrated charge system are the competition between classical phases and the emergence of quantum phases. 
A typical theoretical example is seen in an interacting spinless fermion model on an anisotropic triangular lattice.~\cite{hotta} 
Two-types of the classical charge ordered (CO) phases, termed the ``horizontal-stripe" and ``vertical-stripe" COs, are realized according to the anisotropy of the inter-site Coulomb interactions. Around the frustration point characterized by the isotropic Coulomb interactions, a 3-fold CO metallic state appears owing to the fermion kinetic energy.  
This model has been applied to the CO phenomena in molecular organic solids, and has successfully explained the  structural, transport, and dielectric properties. 

In contrast to the frustrated magnets, charge dynamics beyond the static CO structure in the frustrated charge systems have not been fully touched yet.~\cite{Shingai,Nishimoto,Seo,Cano} 
This might be attributed to the fact that the experimental probes which can access directly to the charge fluctuation in a wide range of energy and momentum are limited.  
Recently, the  resonant and non-resonant inelastic x-ray scattering techniques have been developed as the standard tools  to explore the dynamical charge correlation; they play the same roles with the inelastic neutron scattering technique in frustrated magnets. 

\begin{figure}[t]
\begin{center}
\includegraphics[width=\columnwidth,clip]{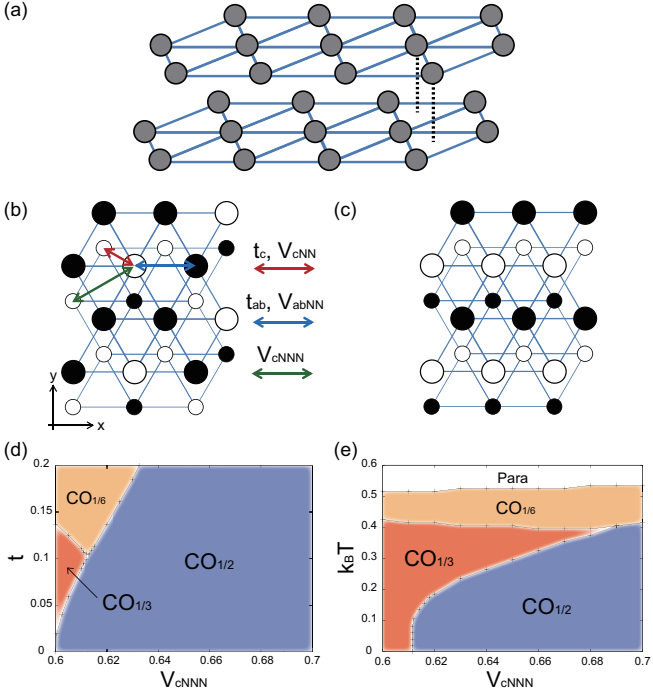}
\end{center}
\caption{(Color online) 
(a) A lattice structure of a paired triangular lattice. 
(b) Schematic top views of the CO$_{1/3}$ and (c) CO$_{1/2}$ structures. 
Filled and open circles represent charge rich and poor sites, respectively, 
and large and small circles represent the sites in the upper and lower planes, respectively. 
The Coulomb interactions and the fermion hoppings introduced in the model Hamiltonian are shown by arrows. 
(d) Phase diagrams on the $V_{\rm cNNN}-t$ plane at $T=0$, and (e) on the $V_{\rm cNNN}-T$ plane at 
$t=0.1$. Abbreviation ``Para" implies a charge disordered phase. 
}
\label{fig:GS}
\end{figure}

In this letter, we study charge dynamics in a frustrated charge system in a wide range of energy, momentum and temperature ($T$). 
As a typical frustrated charge system, we adopt an interacting spinless fermion model on a paired-triangular lattice (see Fig.~\ref{fig:GS}(a)). 
This was proposed as an electronic model in the layered iron oxides, LuFe$_2$O$_4$~\cite{yamada,ikeda,groot}; a candidate material of  the electronic ferroelectricity where a CO induces the electric polarization.~\cite{khomskii,ishihara}  
A CO phase diagram is analogous to that in a spinless fermion model on a triangular lattice;~\cite{nagano,naka1,watanabe1,watanabe2}  two classical COs termed the two-fold CO (abbreviated CO$_{1/2}$) and four-fold CO (CO$_{1/4}$) are caused by the inter-site Coulomb interactions, and between the two COs, the three-fold CO (CO$_{1/3}$) appears owing to the thermal or quantum fluctuations. The inversion symmetry in CO$_{1/3}$ is broken, since charge distributions are different between upper and lower layers. 
The recently observed optical and resonant inelastic x-ray scattering (RIXS) spectra in LuFe$_2$O$_4$ provide a good touchstone of the theoretical calculations based on this model.~\cite{Xu,Lee,itoh,Ishii} 
In particular, we focus on the charge dynamics in the CO$_{1/3}$ and CO$_{1/2}$ phases (see Figs.~\ref{fig:GS}(b) and (c)). 
The optical conductivity spectra in CO$_{1/3}$ show multiple components and their low-energy weights are survived even below the charge order temperature ($T_{\rm CO}$). These are related to the stability of CO$_{1/3}$, and are sharply in contrast to the spectra in CO$_{1/2}$. 
Change of the dynamical charge correlation below $T_{\rm CO}$ is weakly momentum dependent in CO$_{1/3}$, 
while an abrupt reduction is observed in CO$_{1/2}$. 
Results in CO$_{1/3}$ are attributable to its characteristic CO structure realized owing to the frustration effects, 
and explain some aspects of the recent optical and RIXS experiments. 

The model Hamiltonian for interacting fermion system on a paired-triangular lattice is given by~\cite{watanabe1,watanabe2} 
\begin{align}
{\cal H}=-\sum_{(ij)} t_{ij} \left (c^\dagger_i c_j^{}+H. c. \right )
+\sum_{(ij)} V_{ij} n_i n_j , 
\label{eq:vt}
\end{align}
where $c_i^\dagger$ and $c_i$, respectively, are the creation and annihilation operators for a spin-less fermion at site $i$, 
and $n_i=c_i^\dagger c_i^{}$ is a number operator. 
The first and second terms represent the inter-site hoppings and the Coulomb interactions, respectively. 
As shown in Fig.~\ref{fig:GS}(b), we introduce the Coulomb interactions between the inter-layer nearest-neighbor (NN) sites ($V_{\rm cNN}$), the intra-layer NN sites ($V_{\rm abNN}$) and the inter-layer next NN sites ($V_{\rm cNNN}$). 
The transfer integrals are considered between the inter-layer NN sites ($t_{\rm cNN}$) and the intra-layer NN sites ($t_{\rm abNN}$). 

In the numerical calculation, energy parameters are given as a unit of $V_{\rm abNN}$. 
We consider the cases of $V_{\rm cNN}=1.2$, and $t \equiv t_{\rm cNN}=t_{\rm abNN}$. 
We use the cluster mean-field method; 
the Hamiltonian in a finite-size cluster, where  the mean fields are applied at the edge sites, is solved, and the mean fields are obtained self-consistently with the solution inside of the cluster. 
 At the edge sites, the hopping terms in the Hamiltonian are neglected, and the Coulomb interaction terms are decoupled as $V_{ij} n_i n_j  \rightarrow V_{ij}(n_i \langle n_j \rangle+ \langle n_i \rangle n_j-\langle n_i \rangle \langle n_j \rangle)$.
The exact diagonalization method based on the Lanczos (Householder) algorithm is adopted 
in the zero-temperature (finite-temperature) calculations, where the cluster size is taken to be up to $12 \times 2$ sites ($6 \times 2$ sites).

%
We first introduce the static charge structure before showing the charge dynamics~\cite{naka1,watanabe1,watanabe2}. 
The charge structures are identified by the order parameters defined by $\left| \langle n_{\bm q} \rangle \right| = \left| N^{-1}\sum_i \langle n_i \rangle e^{- i {\bm q}\cdot {\bm r_i}} \right|$, where $N$ and ${\bm r}_i$ are the total number of sites and the position at site $i$, respectively. 
The phase diagrams on the planes of $t-V_{\rm cNNN}$ and $T-V_{\rm cNNN}$ are presented in Figs.~\ref{fig:GS}(d) and (e), respectively. 
We focus on the region of $V_{\rm cNNN} > 0.6$, in which 
the two COs, CO$_{1/2}$  and CO$_{1/3}$, 
compete with each other. 
The CO$_{1/2}$ and CO$_{1/3}$ structures are characterized by the momenta ${\bm q}_{1/2}\equiv (-1/2, 1/2)$ and ${\bm q}_{1/3} \equiv (-1/3, 2/3)$, respectively.
%
At $T=0$ (see Fig.~\ref{fig:GS}(d)), the CO$_{1/2}$ phase is stabilized in a wide rage of the $t-V_{\rm CNNN}$ plane, and the CO$_{1/3}$ appears around $V_{\rm cNNN}=0.6$ in finite $t$. 
With increasing $t$, CO$_{1/3}$ overcomes CO$_{1/2}$,  and CO$_{1/6}$ characterized by the momentum $(1/6, 1/6)$ appears. 
A topologically similar phase diagram is obtained on the $T-V_{\rm cNNN}$ plane as shown in Fig.~\ref{fig:GS}(e). 
A finite phase space of CO$_{1/3}$ at $T=0$ is owing to a finite hopping parameter of $t=0.1$. 
%
It is known from the previous calculations at $T=0$ that,  in the region of $V_{\rm cNNN} < 0.6$, another CO characterized by ${\bm q}=(1/4, 1/4)$ plays a similar role with CO$_{1/2}$ for $V_{\rm cNNN} > 0.6$.~\cite{naka1,watanabe1,watanabe2} 
This CO is not reproduced in the present calculations, in which the cluster sizes is limited to be smaller than the previous calculations where the static charge structures were examined. 
Namely, CO$_{1/3}$ is realized by the thermal and quantum charge fluctuations, while CO$_{1/2}$ is stabilized by 
the inter-site Coulomb interactions. 

\begin{figure}[t]
\begin{center}
\includegraphics[width=\columnwidth,clip]{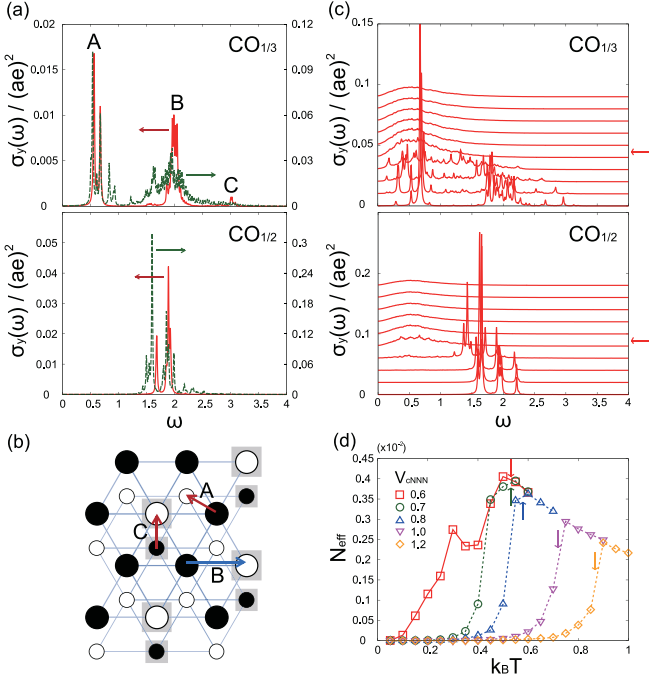}
\end{center}
\caption{(Color online) 
(a) The optical conductivity spectra at $T=0$ in the CO$_{1/3}$ ($V_{\rm cNNN}=0.6$) and CO$_{1/2}$ ($V_{\rm cNNN}=0.7$) phases. 
Solid and dotted lines are for the spectra at $t=0.04$ and $0.13$, respectively. 
Symbols A, B and C correspond to the three kinds of charge excitations shown in (c). 
(b) A schematic CO$_{1/3}$ structure, where strong-potential sites are shaded by gray squares (see the text). 
Arrows represent three kinds of charge excitations. 
(c) The optical conductivity spectra in finite temperatures in the CO$_{1/3}$ ($V_{\rm cNNN}=0.6$) and CO$_{1/2}$ ($V_{\rm cNNN}=0.7$) phases. 
Temperatures are changed from $0.1$ to $1$ from the bottom to the top. 
Arrows indicate $T=T_{\rm CO}$. 
A parameter value is chosen to be $t=0.1$. 
(d) Temperature dependences of the integrated optical conductivity spectra, $N_{\rm eff}$, for several values of $V_{\rm cNNN}$. 
Solid and broken lines are for the results where the ground states are CO$_{1/3}$ and CO$_{1/2}$, respectively. 
Arrows indicate $T_{\rm CO}$s. 
}
\label{fig:sigma}
\end{figure}

Now, we introduce the charge dynamics in the two CO phases, i.e. CO$_{1/2}$ and CO$_{1/3}$. 
The optical conductivity spectra 
at $T=0$ in CO$_{1/3}$ ($V_{\rm cNNN}=0.6$) and in CO$_{1/2}$ ($V_{\rm cNNN}=0.7$)  are presented in Figs.~\ref{fig:sigma}(a) . 
The optical conductivity spectra are defined by 
\begin{align}
\sigma_{ \mu } (\omega)=-\frac{e^2}{N\omega}  {\rm Im} 
\sum_m 
&\biggl ( 
\frac{\langle 0| j^\mu |m \rangle \langle m|j^\mu| 0\rangle}{\omega-E_m+E_0+i \eta}
\nonumber \\
& +
\frac{\langle 0| j^\mu |m \rangle \langle m|j^\mu| 0\rangle}{\omega+E_m-E_0+i \eta}
\biggr ) , 
\end{align}
where $| m \rangle$ and $E_m$ are the $m$-th eigen state and eigen energy, respectively, $j^\mu$ is a current operator with the Cartesian coordinate $\mu$, and $\eta$ is an infinitesimal constant. 
The polarization direction is taken parallel to the $y$ direction, and $\eta$ is chosen to be $0.01$ in the numerical calculations.
Broad spectra in CO$_{1/3}$ spread over a wide energy range.  
As shown in the results with a small hopping integral (see solid lines), 
the spectra are decomposed into the three components, signed A, B, and C. 
In sharp contrast, the spectra in CO$_{1/2}$ are located around $1.5 \lsim \omega \lsim 2$. 
%
A schematic charge configuration and possible charge excitations in CO$_{1/3}$ are shown in Fig.~\ref{fig:sigma}(b). 
Two-inequivalent sites are identified; 
i) charge rich (poor) sites shaded by gray squares which are surrounded by six intra-plane NN poor (rich) sites, 
and ii) the sites surrounded by three intra-plane NN poor and rich sites. 
These are termed the strong- and weak-potential sites, respectively, from now on. 
As a result, three kinds of the charge excitations occur, that is, excitations between the two strong-potential sites, 
the two weak-potential sites, and the strong- and weak-potentials sites termed C, A, and B, respectively.  
The multiple-peak structures in the optical conductivity spectra are attributable to these excitations. 
On the other hand, no inequivalent sites exist in CO$_{1/2}$ as shown in Fig.~\ref{fig:GS}(c). 
The characteristic low-energy excitations denoted as A in Fig.~\ref{fig:sigma}(a) induce large thermal/quantum charge fluctuations in equilibrium states, and are responsible for a stability of CO$_{1/3}$ in finite   
$t$ and $T$ as shown in Figs.~\ref{fig:GS} (d) and (e). 

%
The optical conductivity spectra in CO$_{1/3}$ and CO$_{1/2}$ in finite $T$ are presented in Figs.~\ref{fig:sigma}(c). 
In both the two COs, 
featureless broad peaks centered around $\omega=0.5$ are seen in $T>T_{\rm CO}$. 
Below $T_{\rm CO}$, the spectra in CO$_{1/3}$ spread in a wide energy range 
and show the multiple-peak structure mentioned above. 
In sharp contrast, in CO$_{1/2}$,  the spectral weights are pushed up and a clear CO gap opens below $T_{\rm CO}$. 
We do not find any significant qualitative difference between the optical spectra calculated in $N=12$ and $24$. 
In order to examine the detailed temperature dependence of the optical conductivity spectra, we introduce the integrated spectral intensities, 
$N_{\rm eff}=\int_0^\omega \sigma_{\mu}(\omega') d\omega' $. 
The results for several values of $V_{\rm cNNN}$ are shown in Fig.~\ref{fig:sigma}(d) where $\omega$ is chosen to be $0.5$ which is lower than the optical gap energy $0.6 (=V_{\rm cNNN})$. 
Above $T_{\rm CO}$,  $N_{\rm eff}$ increases with decreasing $T$ for all values of $V_{\rm cNNN}$. 
Increasing and decreasing of $N_{\rm eff}$ above and below $T_{\rm CO}$, respectively, 
reflect a metal-insulator transition and  an opening of the CO gap.
In contrast to an abrupt reduction of $N_{\rm eff}$ in CO$_{1/2}$, 
$N_{\rm eff}$ in CO$_{1/3}$ survives down to far below $T_{\rm CO}$. 
This is owing to the weak-potential sites in CO$_{1/3}$ resulted from the frustration effect. .

\begin{figure}[t]
\begin{center}
\includegraphics[width=\columnwidth,clip]{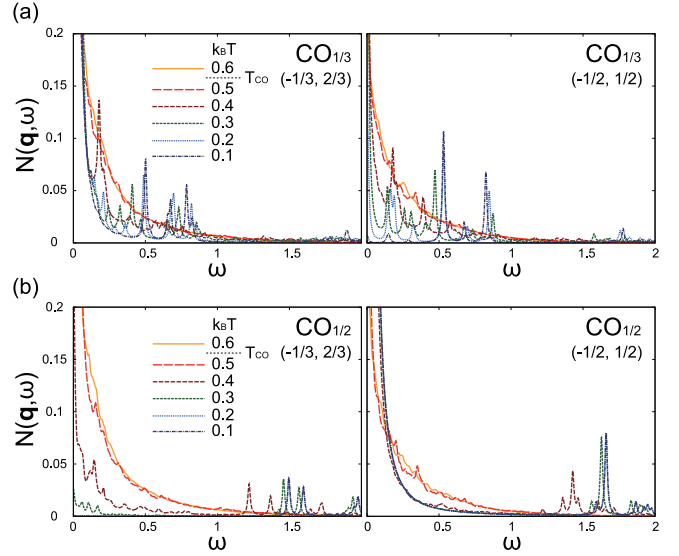}
\end{center}
\caption{(Color online) 
(a) Temperature dependences of the dynamical charge correlation functions in the CO$_{1/3}$ ($V_{\rm cNNN}=0.6$) , and (b) the CO$_{1/2}$ ($V_{\rm cNNN}=0.7$) phases.  
Momenta are given in the Brillouine zone for the triangular lattice. 
Parameters are chosen to be $t=0.1$ and $\eta = 0.01$.    
}
\label{fig:nqwfinitet}
\end{figure}

The momentum dependence of the charge fluctuations is examined by calculating the dynamical charge-correlation function defined as  
\begin{align}
N({\bm q}, \omega) =- {\rm Im}
\sum_m \frac{\langle 0| n_{-{\bm q}}|m \rangle \langle m | n_{\bm q}| 0 \rangle}
{\omega-E_m+E_0+i \eta}. 
\end{align}
%
%
Temperature dependences of the dynamical charge correlation functions in CO$_{1/3}$ and CO$_{1/2}$ are presented in Fig.~\ref{fig:nqwfinitet} at representative momenta of ${\bm q}_{1/3}$ and ${\bm q}_{1/2}$. 
Above $T_{\rm CO}$, broad peaks around $\omega=0$ are only confirmed in both COs. 
These peaks correspond to the quasi-elastic diffusive peaks and do not show remarkable momentum dependences. 
Peak widths seem to be governed by temperature.
Below $T_{\rm CO}$,  the diffusive peaks shrunk into sharp peaks around $\omega=0$ 
at ${\bm q}_{1/3}$ in CO$_{1/3}$, and at ${\bm q}_{1/2}$ in CO$_{1/2}$, reflecting the static charge orderings. 
On the other hand, diffusive peaks disappear and fine structures in finite energies emerge at  ${\bm q}_{1/2}$ in CO$_{1/3}$ and at ${\bm q}_{1/3}$ in CO$_{1/2}$. 

\begin{figure}[t]
\begin{center}
\includegraphics[width=\columnwidth,clip]{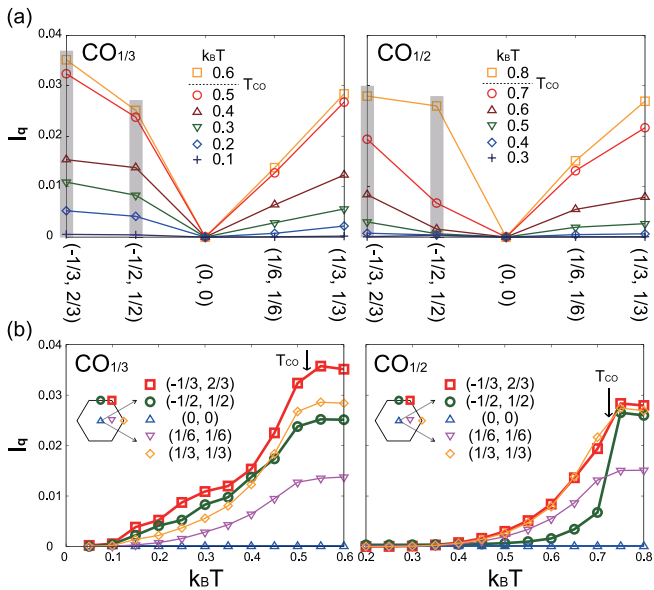}
\end{center}
\vspace{-0.2cm}
\caption{(Color online) 
(a) Momentum dependences of the integrated intensities of the dynamical charge correlation functions for several $T$ in the CO$_{1/3}$ ($V_{\rm cNNN}=0.6$) and CO$_{1/2}$ ($V_{\rm cNNN}=1.0$) phases. 
Momenta are represented in the Brillouin zone for the triangular lattice. 
A parameter value is chosen to be $t=0.1$. 
(b) Temperature dependences of the integrated intensities of the dynamical charge correlation functions in the CO$_{1/3}$ ($V_{\rm cNNN}=0.6$) and CO$_{1/2}$ ($V_{\rm cNNN}=1.0$) phases. 
Arrows represent $T_{\rm CO}$. 
The inset shows the first Brillouin zone and momenta in which the correlation functions are calculated.
} 
\label{fig:nqwfinitet2}
\end{figure}

Detailed temperature dependences of the dynamical charge correlations are studied by calculating the integrated intensity defined by 
$I_{\bm q}=\int_0^\omega  N ({\bm q}, \omega') d \omega'$  
where the sharp peaks centered around zero energy below $T_{\rm CO}$, corresponding to the super-lattice peaks, 
are removed from the integrand, and the upper limit of integral $\omega$ is chosen to be 0.5. 
Results for several $T$ are plotted as functions of ${\bm q}$ in Figs.~\ref{fig:nqwfinitet2}(a). 
In CO$_{1/2}$, abrupt reduction of $I_{\bm q}$ at ${\bm q}_{1/2}$ reflects a growth of the peak around  zero energy below $T_{\rm CO}$. 
Reductions at other ${\bm q}$ are more moderate than  $I_{{\bm q}_{1/2}}$. 
In contrast, no remarkable momentum dependences are confirmed in CO$_{1/3}$; $I_{\bm q}$ at all momenta monotonically decrease with decreasing $T$. 
Difference between the two COs are clearly shown in Figs.~\ref{fig:nqwfinitet2}(b), 
where the integrated intensities for several ${\bm q}$ are replotted as functions of temperature. 
This characteristic momentum dependence in CO$_{1/3}$ is supposed to be due to the charge frustration effect 
which suppresses charge fluctuations at a specific momentum. 
%
%

It is mentioned that the characteristic charge dynamics shown in the present calculations are also expected in other charge frustrated systems. 
A similar multiple-peak structure in the optical conductivity spectra in CO$_{1/3}$ (see Fig.~\ref{fig:sigma}(a)) is confirmed in the metallic 3-fold CO phase, termed ``pinball liquid state", realized in the spinless fermion model on a two-dimensional triangular lattice. 
There, localized charges with the three-fold periodicity termed ``pins" coexist with mobile charges termed ``balls".~\cite{hotta}
The wave function in this CO phase is given by a linear combination of the classical charge configurations, in which 
inequivalent ``pin" sites due to different NN ``ball" configurations exist. 
The observed multiple peak structure in the optical conductivity spectra is attributed to the inequivalent ``pin" sites, in the same manner as the spectra in CO$_{1/3}$ on the paired  triangular lattice.  

Finally, we discuss the relation of the present results to the recent experimental observations in LuFe$_2$O$_4$. The optical conductivity spectra were measured in Ref.~\cite{Xu}, 
where the spectra around 1-1.5eV and above 2eV are assigned as 
the charge excitations between Fe$^{2+}$ and Fe$^{3+}$ and those inside of Fe$^{2+}$, respectively.  
The effective oscillator strength for the Fe$^{2+}$-Fe$^{3+}$ excitations, corresponding to $N_{\rm eff}(\omega)$ in the present calculations, remains even below $T_{\rm CO}$, and is saturated around the magnetic ordering temperature. The authors claims that these data support the order-by-fluctuation mechanism for development of the CO. These observations are consistent with the calculated results in Fig.~\ref{fig:sigma}(d). 
We expect that the spin ordering, which are not taken into the present model, suppresses the charge fluctuation due to the so-called ``gate effect" in the spin-dependent electron hopping, as known in the double-exchange interaction. 
Momentum resolved charge fluctuation in LuFe$_2$O$_4$ were recently observed by RIXS experiments where the incident x-ray is tuned around the Fe $K$-edge.~\cite{Ishii} 
A peak structure found around 1eV was ascribed as the Fe$^{2+}$ and Fe$^{3+}$ charge excitation. 
Monotonic reductions of the peak intensities below $T_{\rm CO}$ were observed at not only the momentum characterizing the long-range CO but also other momenta. 
This is in contrast to the experimental results in other CO systems, in which the charge frustration is not expected,~\cite{Ishii,wakimoto} 
similarly the results shown in Fig.~\ref{fig:nqwfinitet2}. 
%

In summary, we examine charge dynamics in the interacting spinless fermion model on a paired-triangular lattice, as a typical model for the frustrated charge system, in wide ranges of energy, momentum and temperature.  
We focus on the dynamics in the two kinds of COs, CO$_{1/2}$ and CO$_{1/3}$, which are recognized as the classical and quantum COs realized by the Coulomb interaction and the fermion kinetic energy, respectively. 
The optical conductivity spectra in CO$_{1/3}$ show the multiple components where the low energy weights remain to be survive far below $T_{\rm CO}$, in contrast to the spectra in CO$_{1/2}$. 
The dynamical charge correlation functions show also large differences in the two COs; 
reductions in the charge fluctuations below $T_{\rm CO}$ show weak momentum dependence in CO$_{1/3}$, while a rapid decreasing of the fluctuation at ${\bm q}_{1/2}$ is seen in CO$_{1/2}$. 
These differences in the two COs are attributed to the characteristic CO structure in CO$_{1/3}$, 
and reproduce the recent optical and RIXS experiments. 
The present work demonstrate a new route, in which the characteristic charge dynamics in frustrated CO systems are directly monitored by the experiments. 
 
We thank K.~Ishii, N.~Ikeda, S.~Iwai, T.~Watanabe, and J.~Nasu for helpful discussions. 
MN thanks to valuable discussion with H. Seo. 
This work was supported by JSPS KAKENHI Grant Numbers 26287070. Some of the numerical calculations
were performed using the supercomputing facilities at ISSP, the University of Tokyo.




\begin{thebibliography}{99}


\bibitem{jpsj} 
{\it Special Topics: Novel States of Matter Induced by Frustration}, 
J. Phys. Soc. Jpn. {\bf 79}, 001101-001112 (2010).

\bibitem{diep}
{\it Frustrated Spin Systems}, edited by H. T. Diep, (World Scientific, New Jersey, 2004).

\bibitem{poilblanc}
D. Poilblanc, and H. Tsunetsugu, 
{\it in Introduction to Frustrated Magnetism}, 
edited by C. Lacroix, P. Mendels, F. Mila, 
(Springer, Heidelberg, 2011). 

\bibitem{seo} 
H. Seo, J. Merino, H. Yoshioka, and M. Ogata,  J. Phys. Soc. Jpn. {\bf 75}, 051009 (2006). 

\bibitem{merino} 
J. Merino, H. Seo and M. Ogata, Phys. Rev. B {\bf 71}, 125111 (2005).

\bibitem{hotta}
C. Hotta, N. Furukawa, A. Nakagawa, and K. Kubo,
J. Phys. Soc. Jpn. {\bf  75}, 123704 (2006). 


\bibitem{Shingai}
M. Shingai, S.Nishimoto, and Y.Ohta, 
J. Phys. Chem.Sol. {\bf 69}, 3382 (2008). 

\bibitem{Nishimoto}
S. Nishimoto, M. Shingai, and Y. Ohta, 
Phys. Rev. B {\bf 78}, 035113 (2008).

\bibitem{Seo}
H. Seo, K. Tsutsui, M. Ogata and J. Merino, 
J. Phys. Soc. Jpn.  {\bf 75}, 114707 (2006).

\bibitem{Cano}
L. Cano-Cort${\rm \acute{ e}}$s, J. Merino, and S. Fratini, 
Phys. Rev. Lett. {\bf 105}, 036405 (2010). 

\bibitem{yamada} 
Y. Yamada, S. Nohdo, and N. Ikeda, J. Phys. Soc. Jpn.  {\bf 66}, 3733 (1997).

\bibitem{ikeda}
N. Ikeda, H. Ohsumi, K. Ohwada, K. Ishii, T. Inami, K. Kakurai, Y. Murakami, K. Yoshii, S. Mori, Y. Horibe 
and H. Kito, Nature, {\bf 436}, 1136 (2005). 

\bibitem{groot}
J. de Groot, T. Mueller, R. A. Rosenberg, D. J. Keavney, Z. Islam, J.-W. Kim, and M. Angst, 
Phys. Rev. Lett. {\bf 108}, 187601 (2012). 



\bibitem{nagano}
A. Nagano, M. Naka, J. Nasu and S. Ishihara, 
Phys. Rev. Lett. {\bf 99}, 217202 (2007). 

\bibitem{naka1}
M. Naka, A. Nagano and S. Ishihara, 
Phys. Rev. B {\bf 77}, 224441 (2008). 

\bibitem{watanabe1}
T. Watanabe and S. Ishihara, 
J. Phys. Soc. Jpn. {\bf 78},  113702 (2009).

\bibitem{watanabe2}
T. Watanabe and S. Ishihara, 
J. Phys. Soc. Jpn. {\bf 79}, 114714 (2010). 


\bibitem{khomskii}
J.~van den Brink, and D.~I.~Khomskii, 
J.~Phys.: Cond. Mat. {\bf 20}, 434217 (2008). 

\bibitem{ishihara}
S. Ishihara, 
J. Phys. Soc. Jpn. {\bf 79}, 011010 (2010).



\bibitem{Xu}
X. S. Xu, M. Angst, T. V. Brinzari, R. P. Hermann, J. L. Musfeldt, A. D. Christianson, D. Mandrus, B. C. Sales, S. McGill, J.-W. Kim, and Z. Islam, 
Phys. Rev. Lett. {\bf 101}, 227602 (2008). 

\bibitem{Lee}
J. Lee, S. A. Trugman,	C. D. Batista, C. L. Zhang, D. Talbayev, X. S. Xu, S.-W. Cheong, D. A. Yarotski, A. J. Taylor and R. P. Prasankumar, Sci. Rep. {\bf  3}, 2654 (2013). 

\bibitem{itoh}
H. Itoh, K. Itoh, K. Anjyo, H. Nakaya, H. Akahama, D. Ohishi, S. Saito, T. Kambe, S. Ishihara, N. Iked, and S. Iwai, 
J. Luminesc. {\bf 133}, 149 (2013). 

\bibitem{Ishii}
M. Yoshida, K. Ishii, M. Naka, S. Ishihara, I. Jarrige, K. Ikeuchi, Y. Murakami, K. Kudo, Y. Koike, N. Ikeda and J. Mizuki, Meeting abstract of the Phys. Soc. Jpn. {\bf 66}-2, 541 (2011). 

\bibitem{wakimoto}
S. Wakimoto, K. Ishii, H. Kimura, K. Ikeuchi, M. Yoshida, T. Adachi, D. Casa, M. Fujita, Y. Fukunaga, T. Gog, Y. Koike, J. Mizuki, and K. Yamada, 
Phys. Rev. B {\bf 87}, 104511 (2013). 


\end{thebibliography}
\end{document}